\documentclass[review]{elsarticle}
\usepackage{amsmath}
\usepackage{hyperref}
\usepackage{amssymb}
\usepackage{graphicx}%
\newcommand{\be}{\begin{equation}}
\newcommand{\ee}{\end{equation}}
\newcommand{\bea}{\begin{eqnarray}}
\newcommand{\eea}{\end{eqnarray}}

\newcommand{\D}{\partial}

\newcommand{\R}{{\mathbb{R}}}
\begin{document}

\begin{frontmatter}
 \title{Integrable Motion of  Curves in Self-Consistent Potentials : Relation to Spin Systems and Soliton Equations}

\author{R. Myrzakulov} 
\author{G. K. Mamyrbekova} \author{G.N. Nugmanova}\author{K.R.Yesmakhanova}
\address{Eurasian International Center for Theoretical Physics and  Department of General \&  Theoretical Physics, Eurasian National University, Astana 010008, Kazakhstan} 

\author{M. Lakshmanan{\corref{mycorrespondingauthor}}}
\cortext[mycorrespondingauthor]{Corresponding author}
\ead{lakshman@cnld.bdu.ac.in}
\address{Centre for Nonlinear Dynamics, School of Physics, 
Bharathidasan University, Tiruchirapalli 620 024, India}



 \renewcommand{\baselinestretch}{1.1}

 \begin{abstract}
Motion of curves and surfaces in $\R^3$ lead to nonlinear evolution equations which are often integrable. They are also intimately connected to the dynamics of spin chains in the continuum limit and integrable soliton systems through geometric and gauge symmetric connections/equivalence. Here we point out the fact that a more general situation in which the curves evolve in the presence of additional self consistent vector potentials can lead to interesting generalized spin systems with self consistent potentials or soliton equations with self consistent potentials. We obtain the general form of the evolution equations of underlying curves and report specific examples of generalized spin chains and soliton equations. These include principal chiral model and various Myrzakulov spin equations in (1+1) dimensions and their geometrically equivalent generalized nonlinear Schr\"odinger (NLS) family of equations, including Hirota-Maxwell-Bloch equations, all in the presence of self consistent potential fields. The associated gauge equivalent Lax pairs are also presented to confirm their integrability.
 \end{abstract}


\end{frontmatter}

 
\section{Introduction}
Integrable soliton equations have interesting geometric connections/equivalence with moving space curves and surfaces both in (1+1) and (2+1) dimensions[1-6]. These connections especially manifest through integrable spin chains. One of the most interesting connections is the mapping of the Heisenberg spin chain onto the integrable nonlinear Schr\"odinger equation, where the square of curvature of the moving curve is related to the energy density of the spin chain and the torsion is related to the current density[7,8]. Many of the other soliton equations can also be given such connections~[1-3]. This relationship can also be reinterpreted as a gauge transformation between the spin systems and soliton equations so that the Lax pairs between the two systems can be mapped onto each other and so also the zero curvature conditions[9,10].

One can also extend this interconnection to moving curves and surfaces in (2+1) dimensions~[6,11-14]. In this way one can identify topological conserved quantities with geometrical invariants. For example, one can map Ishimori spin equation and Myrzakulov-I equation with Davey-Stewartson and Zakharov-Strachan (2+1) dimensional nonlinear Schr\"odinger equation~[14].

In this paper, we present a further generalization by incorporating an additional self consistent potential in the presence of which the curves and surfaces move. The corresponding generalized evolution equations for the moving curves in $\R^3$  is presented. The generalization to moving surfaces will be presented separately. Several interesting generalized spin systems and soliton equations with self consistent potentials can then be identified. These include the principal chiral field equation, various generalizations of Myrzakulov family of spin equations~[15,16] in (1+1) dimensions and their geometrically equivalent counterparts of generalized nonlinear Schr\"odinger family of equations in the presence of self consistent vector fields.

The plan of the paper is as follows. In Sec.2, we briefly review the nonlinear dynamics of moving space curves in (1+1) dimensions and deduce the evolution equations for the curvature and torsion of the curve. In Sec.3, we generalize the motion equations in the presence of a self consistent vector potential and deduce the modified form of these equations which now also involve the components of the vector potentials. In Sec. 4, we identify several specific cases of spin systems in the presence of vector potentials which can be mapped onto the moving curves. These are then transformed into generalized soliton equations. In Sec.5, the gauge equivalent Lax pairs are presented for the various systems discussed in the previous section to prove the integrability of them. Then in Sec. 6 the equivalent induced surfaces are identified.  Finally in Sec.7, we present our conclusions.
\section{Motion of curves in 1+1 dimensions}

We consider a space curve in $\R^3$. In 1+1 dimensions the motion of such curves is  defined by the following Serret-Frenet equations and rigid body equation [2], respectively,
\begin{eqnarray}
\frac{\partial}{\partial x}
\left ( \begin{array}{ccc}
{\vec  e}_{1} \\
{\vec  e}_{2} \\
{\vec  e}_{3}
\end{array} \right) = C
\left ( \begin{array}{ccc}
{\vec  e}_{1} \\
{\vec  e}_{2} \\
{\vec  e}_{3}
\end{array} \right),\frac{\partial}{\partial t}
\left ( \begin{array}{ccc}
{\vec  e}_{1} \\
{\vec  e}_{2} \\
{\vec  e}_{3}
\end{array} \right) = G
\left ( \begin{array}{ccc}
{\vec  e}_{1} \\
{\vec  e}_{2} \\
{\vec  e}_{3}
\end{array} \right). \label{2.1} 
\end{eqnarray}
Here ${\vec e}_{1}, {\vec e}_{2}$ and ${\vec e}_{3}$ are the unit tangent, normal and binormal vectors, respectively, to the curve and $x$ is the arclength parametrising the curve. The unit tangent vector  ${\vec e}_{1}$ is given by ${\vec e}_{1} = \frac{\partial {\vec  r}}{\partial x}=\frac{1}{\sqrt g}\frac{\partial {\vec  r}}{\partial \theta}$, where $g$ is the metric $g=\frac{\partial {\vec  r}}{\partial \theta}. 
\frac{\partial {\vec  r}}{\partial \theta}$ on the curve such that $x(\theta,t) = 
\int\limits_{0}^{\theta}\sqrt{g(\theta^{\prime}, t)}d\theta^{\prime}$. Here 
$\theta $ defines a smooth curve  and ${\vec  r}(\theta, t)$ is the position 
vector of a point on the curve at time $t$. In (\ref{2.1}) $C$ and $G$ are given by
\begin{eqnarray}
C =
\left ( \begin{array}{ccc}
0   & \kappa     & 0 \\
-\kappa & 0     & \tau  \\
0   & -\tau & 0
\end{array} \right) ,\quad
G =
\left ( \begin{array}{ccc}
0       & \omega_{3}  & -\omega_{2} \\
-\omega_{3} & 0      & \omega_{1} \\
\omega_{2}  & -\omega_{1} & 0
\end{array} \right).\label{2.2} 
\end{eqnarray}
As it is well known, the curvature and torsion of the curve are given  
respectively as
\begin{eqnarray}
\kappa & = & ({\vec e}_{1x}\cdot {\vec e}_{1x})^{1\over 2},\nonumber \\
\tau   & = & \kappa^{-2} {\vec e}_{1}\cdot ({\vec e}_{1x} \wedge {\vec e}_{1xx}) .        \label{2.3}
\end{eqnarray}The compatibility condition of the equations (\ref{2.1}) is
\begin{eqnarray}
C_t - G_x + [C, G] = 0,\label{2.4} 
\end{eqnarray}
 or in terms of elements it reads 
 \begin{eqnarray}
\kappa_t    & = & \omega_{3x} + \tau \omega_2, \label{2.5a} \\ 
\tau_t      & = & \omega_{1x} - \kappa \omega_2, \\ \label{2.6a} 
\omega_{2x} & = & \tau \omega_3-\kappa \omega_1. \label{2.7a} 
\end{eqnarray}

The above formalism allows one to construct the so-called the L-equivalents (Lakshmanan equivalence) of spin systems. Here we present an example  which shows how this formalism works. Considering the Heisenberg ferromagnet equation (HFE)[7],
 \begin{eqnarray}
{\vec S}_{t}+{\vec S}\wedge {\vec S}_{xx}=0,\label{2.8c}
\end{eqnarray}
where ${\vec S}=(S_{1}, S_{2}, S_{3})$ is a unit spin vector so that $S_{1}^2+S_{2}^{2}+S_{3}^{2}=1$,  we assume the identification
\begin{eqnarray}
{\vec e}_{1} \equiv {\vec S}. \label{2.9}
\end{eqnarray}
Then from the HFE  (\ref{2.8c}) it follows that
\begin{eqnarray}
\omega_{1}&=&-\frac{\kappa_{xx}}{\kappa} + \tau^2, \label{4.2}\\
\omega_{2}&=&\kappa_{x},\label{4.3}\\
\omega_{3}&=&\kappa\tau.\label{4.4}
\end{eqnarray}
 Substituting these expressions of $\omega_j$  into the equations (\ref{2.5a})-(\ref{2.6a}), we arrive at the system
\begin{eqnarray}
\kappa_{t}-(\kappa\tau)_{x}-\kappa_{x}\tau&=&0, \label{4.8}\\
\tau_{t}+\left(\frac{\kappa_{xx}}{\kappa}\right)_{x}-2\tau\tau_{x}+ \kappa\kappa_{x}&=&0.\label{4.9}
\end{eqnarray}
Let us introduce the following  complex function $q=\frac{\kappa}{2}e^{-i\partial^{-1}_{x}\tau}$.  It is easy to check  that this function satisfies the well known nonlinear Schr\"odinger equation [7]
\begin{eqnarray}
iq_{t}+q_{xx}+2|q|^2q=0.  \label{4.17}
\end{eqnarray}
Thus the HFE is L-equivalent to the NSE and vice versa. Our aim in this paper is to construct L-equivalent counterparts of some integrable spin systems with self-consistent potentials in (1+1) dimensions. 
\section{Moving curves in (1+1) dimensions in the presence of self consistent potentials}

We now introduce a self consistent vector potential $\vec W(x,t)$ in $\R^3$ derivable as 
\be
\frac{\D \vec W}{\D x} = \vec W_x = 2a \vec W \times \vec e_1,
\ee
where $\vec e_1$ is the unit tangent vector, see Sec. 2 and $a$ is a constant parameter.

Expressing $\vec W$ in the basis of the unit orthonormal triad specifying the moving curve as 
\be
\vec W = W_1(x,t) \vec e_1 +W_2(x,t) \vec e_2 + W_3(x,t) \vec e_3,
\ee
the defining equations for the components of the vector potential can be rewritten, after using (1) and (2), as 
\begin{eqnarray}
W_{1x}&=&\kappa W_{2}, \\
W_{2x}&=&-\kappa W_{1}+\tau W_{3}+ 2a W_{3}, \\
W_{3x}&=&-\tau W_{2}-2a W_{2} .
\end{eqnarray}
Note that the above three equations imply $\vec W^2 = W_1^2 + W_2^2 +W_3^2 = C(t)$, where $C(t)$ is a function of $t$ only.

In the presence of the potential field, the evolution equation for the moving curve gets modified due to the self consistent interaction. The underlying evolution equations for the triad can be identified as follows.

The evolution for the unit tangent vector can be modified in a self consistent way as

\be
\vec e_{1t}= \vec \Omega \times \vec e_1 + 2a^{-1} \vec W \times \vec e_1
\ee
where (from (1) and (2))
\be
\vec \Omega = \sum_{i=1}^3 \omega_i \vec e_i, ~~~~ \vec W = \sum_{i=1}^3 W_i \vec e_i.
\ee
Now using the Serret-Frenet equations for the spatial variation of the trihedral along the arc length, see Eq.(1), one can obtain
\be
\vec e_{2t} = \vec \Omega \times \vec e_2 + 2a^{-1}(-W_3\vec e_1 + (W_1 - 2 a \frac{W_3}{\kappa}) \vec e_3)
\ee
and
\be
\vec e_{3t} = \vec \Omega \times \vec e_3 +2a^{-1}(W_2\vec e_1 - (W_1 - 2 a \frac{W_3}{\kappa}) \vec e_2).
\ee
In other words, the dynamical equations specifying the unit trihedral gets modified from (1) in the presence of the self consistent potential $\vec W(x,t)$ as
\begin{eqnarray}
\frac{\partial}{\partial x} \left ( \begin{array}{ccc}
{\vec  e}_{1} \\
{\vec  e}_{2} \\
{\vec  e}_{3}
\end{array} \right) = C
\left ( \begin{array}{ccc}
{\vec  e}_{1} \\
{\vec  e}_{2} \\
{\vec  e}_{3}
\end{array} \right), 
\frac{\partial}{\partial t}\left ( \begin{array}{ccc}
{\vec  e}_{1} \\
{\vec  e}_{2} \\
{\vec  e}_{3}
\end{array} \right) = \widetilde{G}
\left ( \begin{array}{ccc}
{\vec  e}_{1} \\
{\vec  e}_{2} \\
{\vec  e}_{3}
\end{array} \right), \label{2.1a} 
\end{eqnarray}
where
\begin{eqnarray}
C =
\left ( \begin{array}{ccc}
0   & \kappa     & 0 \\
-\kappa & 0     & \tau  \\
0   & -\tau & 0
\end{array} \right) ,\quad
\widetilde{G} =
\left ( \begin{array}{ccc}
0       &\widetilde \omega_{3}  & -\widetilde\omega_{2} \\
-\widetilde\omega_{3} & 0      & \widetilde\omega_{1} \\
\widetilde\omega_{2}  & -\widetilde\omega_{1} & 0
\end{array} \right) =
\left ( \begin{array}{ccc}
0       & \omega_{3}  & -\omega_{2} \\
-\omega_{3} & 0      & \omega_{1} \\
\omega_{2}  & -\omega_{1} & 0
\end{array} \right) + \nonumber \\
2a^{-1}\left ( \begin{array}{ccc}
0       &  W_{3}  & - W_{2} \\
- W_{3} & 0      &  W_1 -2 a \frac{ W_{3}}{\kappa} \\
 W_{2}  & -( W_{1} -2 a \frac{ W_{3}}{\kappa}) & 0
\end{array} \right) .\label{2.2} 
\end{eqnarray}
Note that 
\begin{eqnarray}
\widetilde\omega_{1}&=&\omega_1 + 2a^{-1}( W_1 - 2 a \frac{ W_3}{\kappa}), \nonumber \\
\widetilde\omega_{2}&=&\omega_2 + 2a^{-1} W_2,\label{4.3}\\
\widetilde\omega_{3}&=&\omega_3+2a^{-1} W_3 ,\nonumber
\end{eqnarray}
where again the curvature $\kappa$ and torsion $\tau$ are given by
\be
\kappa  =  ({\vec e}_{1x}\cdot {\vec e}_{1x})^{1\over 2}, ~~~
\tau   = \kappa^{-2} {\vec e}_{1}\cdot ({\vec e}_{1x} \wedge {\vec e}_{1xx}) .        \label{2.3}
\ee
Then the associated compatibility conditions lead to the evolution equations,

 \be
\kappa_t    = (\omega_{3x} + \tau \omega_2) + 2a^{-1}( W_{3x} + \tau  W_2), \label{2.5} 
\ee
\be
\tau_t =  (\omega_{1x} - \kappa \omega_2) + 2a^{-1}\left[ W_{1x}- 2 a \left(\frac{ W_3}{\kappa}\right)_x - \kappa  W_2\right], \label{2.6} 
\ee
and the consistency condition takes the form
\be
\omega_{2x} +2a^{-1} W_{2x}   =  \tau (\omega_3 +2a^{-1} W_{3})-\kappa \left[\omega_1 +2a^{-1}(  W_1 - 2 a \frac{ W_3}{\kappa})\right] \label{2.7} 
\ee
Now using the defining equations (18)-(20) for the vector potential $\vec  W$ on the right hand sides of (29) - (31), we obtain the modified form of the above equations as

 \be
\kappa_t    = (\omega_{3x} + \tau \omega_2) -4  W_2, \label{2.5} 
\ee
\be
\tau_t =  \omega_{1x} - \kappa \omega_2 - 4 (\frac{ W_3}{\kappa})_x, \label{2.6} 
\ee
while the consistency condition remains unchanged as before :
\be
\omega_{2x}  =  \tau \omega_3-\kappa \omega_1. \label{2.7} 
\ee
These three equations have to be augmented by the defining equations for the vector potential $\vec  W = ( W_1, W_2,  W_3)$ given by Eqs. (18) - (20). Thus Eqs. (32) - (34) along with (18)-(20) constitute the dynamical equations defining the motion of the space curves in $\R^3$ in the presence of the self consistent vector potential $\vec W(x,t)$.

Further, it is also of interest to note that the above formalism can also be identified with the motion of a vortex filament~[17] in the presence  of a vector potential or the motion of a flexible string~[18] in such a potential field. Let $\vec r(x,t)$ defines the filament or string specified by
\be
\vec r_t = \alpha \vec e_1 + \beta \vec e_2 + \gamma \vec e_3 + 2 a^{-1}.\vec W, \label{3.22a}
\ee
where $\alpha(x,t), \beta(x,t), \gamma(x,t)  $ are given functions of x and t. Then taking an $x$ derivative of (\ref{3.22a}), we can map it on the above moving curve equations (25), giving rise to useful physical situations.  One can consider even more general potential fields than (16), namely
\be
\frac{\partial W}{\partial x}=\vec W \times [a_1(x,t) \vec e_1 + a_2(x,t)\vec e_2 + a_3(x,t)\vec e_3]
\ee
and similarly generalize Eq.~(21) as 
\be
\vec e_{1t}=\vec \Omega \times \vec e_1 + \vec W \times [b_1(x,t)\vec e_1 + b_2(x,t)\vec e_2 + b_3(x,t) \vec e_3]
\ee
and study the consequences.  This will be reported separately.

\section{Spin equations, moving space curves and generalized NLS equations}
In view of the above mentioned natural connection between the moving space curve and the one dimensional isotropic Heisenberg spin chain in its continuum limit, we now wish extend this connection to more general situations in the presence of the self consistent vector potentials.  For this purpose we now consider specific evolution equations of spin vectors in the presence of potentials and identify them with moving space curves and then deduce equivalent generalized nonlinear Schr\"odinger family of equations using the formalism presented in the previous section. For this purpose, we map the unit spin vector defining the spin equations with the unit tangent vector of the trihedral associated with the moving space curves. This allows one to make appropriate mapping.

\subsection{The principal chiral equation}
We consider the dynamical equations for the principal chiral field $\vec S(x,t)$ as 
\begin{eqnarray}
{\vec S}_{t}+\frac{2}{a}{\vec S}\wedge {\vec  W}&=&0,\label{3.1}\\
 {\vec W}_{x}+2 a {\vec S}\wedge{\vec  W}&=&0,\label{3.2}
\end{eqnarray}
where $\vec S = (S_1, S_2, S_3)$, $\vec S^2=1$, ${\vec W}=(W_1, W_2, W_3)$.

Introducing the identification 
\begin{eqnarray}
{\vec e}_{1} \equiv {\vec S},\quad {\vec W}=W_1{\vec e}_{1}+W_2{\vec e}_{2}+W_3{\vec e}_{3} \label{3.3}
\end{eqnarray}
with the space curve in $\R^3$, in view of the discussions in Sec. 3, we find 
\begin{eqnarray}
\widetilde\omega_{1}&=& -\frac{4}{\kappa}W_{3}+\frac{2}{\omega}W_{1}, \label{4.2}\\
\widetilde\omega_{2}&=&\frac{2}{\omega}W_{2},\label{4.3}\\
\widetilde\omega_{3}&=&\frac{2}{\omega}W_{3},\label{4.4}
\end{eqnarray}
where $\vec W$ satisfies (18)-(20). Correspondingly the evolution equation for the space curve is given in terms of the curvature $\kappa$ and torsion $\tau$ as 
\begin{eqnarray}
\kappa_{t}+4W_{2}&=&0, \label{4.8}\\
\tau_{t}+4\left(\frac{ W_{3}}{\kappa}\right)_{x}&=&0,\label{4.9}\\
W_{1x}-kW_{2}&=&0, \label{4.10}\\
W_{2x}+kW_{1}-\tau W_{3}-2a W_{3}&=&0, \label{4.11}\\
W_{3x}+\tau W_{2}+2a W_{2}&=&0. \label{4.12}
\end{eqnarray}
We now introduce the following two complex transformations and a redefinition of the real function $W_1$,
\begin{eqnarray}
q(x,t)&=&\frac{1}{2}\kappa(x,t) e^{-i\int_{-\infty}^{x}\tau(x',t)dx'}, \label{4.14}\\
p&=&(iW_{3}-W_{2})e^{-i\int_{-\infty}^{x}\tau(x',t)dx'}, \label{4.15}\\
\eta&=&W_{1}. \label{4.16}
\end{eqnarray}
so that the system (45) - (49) becomes the coupled NLEEs,
\begin{eqnarray}
iq_{t}-2ip&=&0, \label{4.17}\\
p_{x}-2ia p -2\eta q&=&0,\label{4.18}\\
\eta_{x}+q^{*} p +p^{*} q&=&0. \label{4.19}
\end{eqnarray}
The Lax pairs associated with the spin system (\ref{3.1})-(\ref{3.2}) and the NLEEs (52)-(54) can be obtained as the limiting form of the more general systems considered in Sec. 5 which prove the integrability of both the above spin system and the coupled NLEEs.

\subsection{The Heisenberg spin system in the presence of self consistent potential: the M-XCIX   equation}
We consider the generalized Heisenberg spin system in the presence of the vector potential described by the Myrzakulov-XCIX  equation (M-XCIX equation)  [15,16, 21], 
\begin{eqnarray}
\vec{S}_{t}+\vec{S}\wedge \vec {S}_{xx}+\frac{2}{a}\vec{S}\wedge \vec{W}&=&0,\label{3.19}\\
 {\vec W}_{x}+2 a \vec{S}\wedge\vec{W}&=&0,\label{3.20}
\end{eqnarray}
where $\vec{S}$ and $\vec{W}$ are as defined above . Then as before with the identification 
\begin{eqnarray}
\vec{e}_{1} \equiv \vec{S},\quad \vec{W}=W_1\vec{e}_{1}+W_2\vec{e}_{2}+W_3\vec{e}_{3}, \label{4.1}
\end{eqnarray} 
one obtains 
\begin{eqnarray}
\widetilde{\omega_{1}}&=&-\frac{\kappa_{xx}}{\kappa} + \tau^2  -\frac{4}{\kappa}W_{3}+\frac{2}{a}W_{1}, \label{4.2}\\
\widetilde{\omega_{2}}&=&\kappa_{x}+\frac{2}{a}W_{2},\label{4.3}\\
\widetilde{\omega_{3}}&=&\kappa \tau + \frac{2}{a}W_{3},\label{4.4}
\end{eqnarray}
along with (18) - (20). Consequently we obtain the evolution equation for the curvature and torsion as 
\begin{eqnarray}
\kappa_{t}-(\kappa\tau)_{x}-\kappa_{x}\tau+4W_{2}&=&0, \label{4.8}\\
\tau_{t}+\left(\frac{\kappa_{xx}+4 W_{3}}{\kappa}\right)_{x}-2\tau\tau_{x}+ \kappa \kappa_{x}&=&0,\label{4.9}\\
W_{1x}-\kappa W_{2}&=&0, \label{4.10}\\
W_{2x}+\kappa W_{1}-\tau W_{3}-2a W_{3}&=&0, \label{4.11}\\
W_{3x}+\tau W_{2}+2a W_{2}&=&0. \label{4.12}
\end{eqnarray}
The system (61) - (65) gets transformed to the following NLEEs through the transformations (49)-(51) as 
\begin{eqnarray}
iq_{t}+q_{xx}+2|q|^2q-2ip&=&0, \label{4.17}\\
p_{x}-2ia p -2\eta q&=&0,\label{4.18}\\
\eta_{x}+q^{*} p +p^{*} q&=&0, \label{4.19}
\end{eqnarray}
which is a generalized version of the NLS equation. It is also designated as nonlinear Schr\"odinger-Maxwell-Bloch (SMB) equation. It is known to describe optical soliton propagation in fibers with resonant and erbium doped systems [19,20]. The associated Lax pairs for the above systems are also indicated in Sec. 5.

\subsection{Third order generalization of the Heisenberg spin chain  in self consistent potential: the M-XCIV equation}
We consider the following third order generalization of the Heisenberg spin equation in the presence of the vector potential specified by the Myrzakulov-XCIV (M-XCIV) equation
\begin{eqnarray}
{\vec S}_{t}+ \epsilon_1 {\vec S}\wedge {\vec S}_{xx}+\epsilon_2[{\vec S}_{xxx}+6(\beta{\vec S})_{x}]+\frac{2}{\omega}{\vec S}\wedge {\vec  W}&=&0,\label{5.9}\\
 {\vec W}_{x}+2\omega {\vec S}\wedge{\vec  W}&=&0,\label{5.10}
\end{eqnarray}
where $\beta=\frac{1}{8}(\vec S_x\cdot\vec S_x)$, and $\epsilon_1$ and $\epsilon_2$ are contants. Then following the above analysis we can write
\begin{eqnarray}
\widetilde\omega_{1}&=&-\frac{\epsilon_{1}\kappa_{xx}+3\epsilon_{2}(\kappa_{xx}\tau+\kappa_{x}\tau_{x})-4W_{3}}{\kappa} + \epsilon_{1}\tau^2-\epsilon_{2}(\tau_{xx}-\frac{1}{2}\kappa^{2}\tau-\tau^{3})+\frac{2}{a}W_{1},~~~~~~~ \label{6.9}\\
\widetilde\omega_{2}&=&\epsilon_{1}\kappa_{x}+\epsilon_{2}(2\kappa_{x}\tau+\kappa\tau_{x})+\frac{2}{a}W_{2},\label{6.10}\\
\widetilde\omega_{3}&=&\epsilon_{1}\kappa\tau-\epsilon_{2}(\kappa_{xx}-\kappa\tau^{2}+\frac{1}{2}\kappa^{3})+\frac{2}{a}W_{3}. \label{6.11}
\end{eqnarray}
Consequently, we deduce the evolution equations,
\begin{eqnarray}
\kappa_{t}-\epsilon_{1}[(\kappa\tau)_{x}+\kappa_{x}\tau]+\epsilon_{2}[\kappa_{xxx}-3\kappa\tau\tau_{x}-3\kappa_{x}\tau^{2}+\frac{3}{2}\kappa^{2}\kappa_{x}]+4W_{2}&=&0,~~~~~~ \label{6.16}\\
\tau_{t}+(\epsilon_{1}F_{1}+\epsilon_{2}F_{2}+4\frac{W_{3}}{\kappa})_{x}&=&0,\label{6.17}\\
W_{1x}-\kappa W_{2}&=&0, \label{6.18}\\
W_{2x}+\kappa W_{1}-\tau W_{3}-2a W_{3}&=&0, \label{6.19}\\
W_{3x}+\tau W_{2}+2a W_{2}&=&0, \label{6.20}
\end{eqnarray}
where
\begin{eqnarray}
F_{1}=\frac{\kappa_{xx}}{\kappa}-\tau^{2}+\frac{1}{2}\kappa^{2}, \quad F_{2}=\frac{3\kappa_{xx}\tau+3\kappa_{x}\tau_{x}}{\kappa}+\tau_{xx}-\tau^{3}+\frac{3}{2}\kappa^{2}\tau. \label{6.21}
\end{eqnarray}
The above system can in turn be transformed to the Hirota-Maxwell-Bloch equation [19,20],
\begin{eqnarray}
iq_{t}+\epsilon_1(q_{xx}+2|q|^2q)+i\epsilon_2(q_{xxx}+6|q|^2q_x)-2ip&=&0, \label{z20}\\
p_{x}-2i\omega p -2\eta q&=&0,\label{z21}\\
\eta_{x}+q^{*} p +p^{*} q&=&0,\label{z22}
 \end{eqnarray}
by making use of the transformations~(49)-(51).

We can next show that all the above spin equations and associated NLS family of equations are integrable. For this purpose, we report the associated Lax pairs in the next section.

\section{The Lax pairs}
The M-XCIV equation (69)-(70), can be written in the matrix form
\begin{eqnarray}
iS_{t}+\frac{1}{2}\epsilon_1[S, S_{xx}]+i\epsilon_2[ S_{xxx}+6(\beta  S)_{x}]+\frac{1}{a}[S, W]&=&0,\label{6.1}\\
 iW_{x}+a [S, W]&=&0,\label{6.2} 
\end{eqnarray} 
where $S=\vec S \cdot \vec \sigma$ and $W=\vec W \cdot \vec \sigma$, $\vec \sigma = (\sigma_x, \sigma_y,\sigma_z)$ are Pauli matrices. Then the Lax pair associated with (\ref{6.1})-(\ref{6.2}) can be given in a generalized form compared to that of the Lax pair for the simple HFE (7) given by Takhtajan [9],
\begin{eqnarray}
\Phi_{x}&=&U\Phi,\label{6.3}\\
\Phi_{t}&=&(2 \lambda  U+V)\Phi, \label{6.4} 
\end{eqnarray}  
where
 \begin{eqnarray}
U&=&-i  \lambda  S,\label{6.5}\\
V&=&  \lambda ^3V_3+  \lambda ^2V_2+  \lambda  V_{1}+\frac{i}{  \lambda +a}V_{-1}-\frac{i}{a}V_{-1}. \label{6.6} 
\end{eqnarray} 
Here
\begin{eqnarray}
V_3&=&-4i\epsilon_2 S,\label{6.7}\\
V_2&=&-2i\epsilon_1 S+2\epsilon_2SS_x,\label{6.8} \\
V_1&=&\epsilon_1SS_x+\epsilon_2i( S_{xx}+6\beta  S),\label{6.9} \\
V_{-1}&=&W=\begin{pmatrix} W_3&W^{+}\\W^{-}& -W_3\end{pmatrix}, ~~W^{\pm}=W_1 \pm i W_2. \label{6.10} 
\end{eqnarray} 
The corresponding gauge equivalent Lax pair for the Hirota-Maxwell-Bloch equation becomes
 \begin{eqnarray}
\Psi_{x}&=&A\Psi,\label{z7}\\
\Psi_{t}&=&[-4i\epsilon_2  \lambda ^3\sigma_3+B]\Psi,\label{z8} 
\end{eqnarray}  
where 
 \begin{eqnarray}
A&=&-i\lambda \sigma_3+A_0,\label{z9}\\
B&=&\lambda^2B_2+\lambda B_1+B_0+\frac{i}{\lambda+a}B_{-1}.\label{z10} 
\end{eqnarray} 
Here
\begin{eqnarray}
B_2&=&-2i\epsilon_1\sigma_3+4\epsilon_2A_{0},\label{z11}\\
B_1&=&2i\epsilon_2|q|^2\sigma_3+2i\epsilon_2\sigma_3A_{0x}+2\epsilon_1A_0,\label{z11}\\
A_0&=&\begin{pmatrix} 0&q\\-q^{*}& 0\end{pmatrix},\label{z12}\\
B_0&=&(i\epsilon_1|q|^{2}+\epsilon_2(q^{*}_{x}q-q^{*}q_{x}))\sigma_3+B_{01}, \\
B_{01}&=&\begin{pmatrix} 0&i\epsilon_1q_x-\epsilon_2q_{xx}-2\epsilon_2q^{*}q^2\\i\epsilon_1q^{*}_x+\epsilon_2q^{*}_{xx}+2\epsilon_2qq^{*2}& 0\end{pmatrix},\label{z13}\\
B_{-1}&=&\begin{pmatrix} \eta&-p\\-p^{*}& -\eta\end{pmatrix}.\label{z14} 
\end{eqnarray}
The Lax pairs of the other cases in Sec.4 can be obtained by taking appropriate limits, say $\epsilon_2=0$ and so on. The existence of the Lax pairs confirm the integrability of both the spin systems and their generalized NLS family counterparts.  Note that we have not pursued the nature of soliton solutions on integrabity further in this paper, which will be matter of a separate study.
 \section{Integrable surfaces induced by spin systems with self-consistent potentials}
  In this section, as a further consequence of the geometrical aspects, we shortly present main points on the relation between surface geometry and spin systems in (1+1) dimensions with self consistent potentials studied here. As example of spin systems, here we consider the M-XCIX equation (55)-(56) which is integrable. Consider a two-dimensional surface with the position vector ${\vec r}$. Let us now assume that
 \begin{eqnarray}
{\vec r}_{x}\equiv {\vec S},\label{8.1}
\end{eqnarray}
 so that
 \begin{eqnarray}
{\vec r}_{x}^2=1. \label{8.2}
\end{eqnarray}
 The 2-dimensional metric has the form
 \begin{eqnarray}
ds^2=dx^2+2{\vec r}_{x}\cdot {\vec r}_{t}dxdt+{\vec r}_{t}^2dt^2.  \label{8.3}
\end{eqnarray}

In terms of ${\vec r}$ the M-XCIX equation (55)-(56) takes the form
\begin{eqnarray}
{\vec r}_{t}+{\vec r}_{x}\wedge {\vec r}_{xx}-\frac{1}{a^{2}} {\vec W}&=&0, \label{8.4}\\
 {\vec W}_{x}+2a {\vec r}_{x}\wedge{\vec  W}&=&0.  \label{8.5}
\end{eqnarray}
This is the ${\vec r}$ - form of the M-XCIX equation. For the M-XCIX equation (55)-(56) we have
\begin{eqnarray}
{\vec r}_{t}\cdot {\vec r}_{x}&=&\frac{1}{a^{2}} {\vec r}_{x}\cdot {\vec W},\label{8.6}\\
 {\vec r}_{t}^2&=&\frac{C}{a^{4}}-\frac{2}{a^{2}}{\vec W}\cdot ({\vec r}_{x}\wedge {\vec r}_{xx})+ {\vec r}_{xx}^{2}, \label{8.7}
\end{eqnarray}
where $C=C(t)={\vec W}^{2}$. So for the M-XCIX  equation the induced metric of the 2-dimensional  surface is given by
 \begin{eqnarray}
ds^2=dx^2+\frac{2}{a^{2}} {\vec r}_{x}\cdot {\vec W}dxdt+\left[\frac{C}{a^{4}}-\frac{2}{a^{2}}{\vec W}\cdot ({\vec r}_{x}\wedge {\vec r}_{xx})+ {\vec r}_{xx}^{2}\right]dt^2.  \label{8.8}
\end{eqnarray}
This metric defines some 2-dimensional surfaces in $R^{3}$. Having the metric we can calculate all ingredients of the surface. In terms of the $\kappa$ (curvature), $\tau$ (torsion) and functions $W_{j}$,  this metric can be rewritten as
\begin{eqnarray}
ds^2=dx^2+\frac{2}{a^{2}} W_{1}dxdt+\left[\frac{C}{a^{4}}-\frac{2}{a^{2}}\kappa W_{3}+ \kappa^{2}\right]dt^2.  \label{8.9}
\end{eqnarray}
Similar interpretation can be given for other examples also.

\section{Conclusions}
In this paper, we have shown that the nonlinear evolution equations describing the motion of space curves in $R^3$ (equivalently that of the motion of a vortex filament) can be deduced in the presence of self consistent external vector potential fields.  Then mapping onto suitable spin systems, one can identify a general class of integrable nonlinear Schr\"odinger family of equations with additional potential fields.  Our examples include the evolution equations of a principal chiral field, Heisenberg spin and generalized Heisenberg spin chain all in the presence of external potentials.  By making suitable complex transformations, we also map the spin chains/moving curves onto the integrable generalized system of nonlinear nonlinear Schr\"odinger equations, including Hirota-Maxwell-Bloch equations.  The integrability of these systems is also established by deducing the associated gauge connected Lax pairs.  Finally we also pointed out that one can identify the equivalent integrable surfaces induced by the spin systems with self consistent potentials.  As pointed out in the text, these studies can be extended to incorporate even more general class of potentials and also to consider the motion of surfaces.

\section*{Acknowledgments}
The work of M.L. has been supported by a Department of Science and Technology, Government of India sponsored IRHPA research project. M.L. has also been supported by a DST Ramanna Fellowship program research grant and a DAE Raja Ramanna Fellowship.


\end{document}